\documentclass[epj]{svjour}
\usepackage{graphics}
\begin{document}
\title{One more ingredient for energy loss quantification}
\author{ A. M. Hamed\thanks{\email{ahamed@tamu.edu}}
for the STAR Collaboration
%\thanks is optional - remove next line if not needed
%\thanks{\email{ahamed@tamu.edu}}
}                     % Do not remove
\institute{Texas A$\&$M University, Physics Department, Cyclotron Institute, College Station, TX 77843, USA}
\date{Received: date / Revised version: date}
\onecolumn
\abstract{The 
recent results at RHIC for direct $\gamma$-charged hadron azimuthal correlations in heavy-ion collisions are presented. We use these correlations
to study the color charge density of the medium through the medium-induced modification of high-p$_T$ parton fragmentation.
Azimuthal correlations of direct photons at high transverse energy (8 $<$ E$_T$ $<$ 16 GeV) 
with away-side charged hadrons of transverse momentum (3 $<$ p$_T$ $<$ 6 GeV/c) have been measured over a broad range of centrality for 
$Au+Au$ collisions and $p+p$ collisions at $\sqrt{s_{NN}}$ = 200 GeV in the STAR experiment. 
A transverse shower shape analysis in the STAR Barrel Electromagnetic Calorimeter Shower Maximum Detector is used to discriminate between the 
direct photons and photons from the decays of high-p$_T$ $\pi^{0}$. 
The per-trigger away-side yield of direct $\gamma$ is smaller than from $\pi^{0}$ triggers in the same centrality class. 
Within the current uncertainty the recoil suppression in central $Au+Au$ collisions I$_{CP}$ of direct $\gamma$ and $\pi^{0}$ are similar.
 % end of PACS codes
} %end of abstract
\titlerunning{Direct $\gamma$-Jet Measurements}
\authorrunning{A. M. Hamed}
\maketitle
\section{Introduction}
\label{intro}
High-$p_{T}$ particles produced in high energy collisions are of special importance in QCD. The production of these particles arises 
from the hard scattering of the incoming partons and their subsequent fragmentation. 
Therefore, their production rate can be calculated perturbatively, 
and their spectra reveal information about the parent parton distributions in momentum space. 
Furthermore in heavy-ion collisions these particles are formed at the early time of the collisions (``prompt
production") and therefore represent an ideal tool for probing the medium resulted from such collisions. The absorption of the high-$p_{T}$
particles in the medium can be used to obtain a tomographic image of the color charge density of the medium [1].  

Bearing this in mind, the high-$p_{T}$ measurements in
the heavy-ion program show some of the most important results at RHIC: 
The strong suppression of high-$p_{T}$ particles which is observed in central
$Au+Au$ collisions via the dramatic suppression of particle production, and the modification of jet correlations at high transverse momentum
 [2,3,4]. This
strong suppression provides compelling evidence for large energy loss of scattered partons traversing matter that has a high density of color
charges. 

After various complementary measurements from RHIC data 
have revealed the formation of a strongly coupled medium [5,6,7,8,9],
the primary goal of the RHIC heavy-ion program progresses from qualitative statements to rigorous quantitative conclusions. One of the most
important requirements
for quantitative conclusions is the precise measurements of the quantity of the medium-induced energy loss. 

While the single-particle spectra do not provide enough sensitivity to discriminate 
between the different energy loss mechanisms [10,11], the di-hadron azimuthal correlation measurement is expected to
provide somewhat better constraints on 
the energy loss. The initial parton energy is not accessible in single-particle spectra but is somewhat accessible 
in the di-hadron measurements. However a model dependent study [12] has shown that 
at initial high color charge density both single particle spectra and di-hadron azimuthal correlation measurement have diminished sensitivity due to the geometrical bias. 

The $\gamma$-hadron azimuthal correlation measurement has been suggested as a powerful tool to 
quantify the energy loss [13]. In the dominant QCD process of Compton-like scattering, the photon transverse momentum balances 
the parton initial transverse energy. 
In addition, due to the large mean free path of the photon compared to the system size formed in heavy-ion collision, the
direct photon measurement doesn't suffer from the same geometrical bias of that of single particle spectra and di-hadron 
azimuthal correlation measurements.
In particular the $\gamma$-hadron azimuthal correlations provide a unique way to quantify the energy loss 
dependence on the initial parton energy and possibly the color factor. Therefore combining the energy loss measurements from 
many probes of different geometrical biases and different coupling to the formed medium and comparing these measurements with different theoretical
models can lead to a successful quantitative interpretation of the heavy-ion data. 

\section{Data Analysis}
\label{data analysis}
The STAR experiment collected an integrated
luminosity of 535 $\mu {b}^{-1}$ of $Au+Au$ collisions at $\sqrt{s_{NN}}$ = 200 GeV in 2007 using level-2 high-$p_T$ tower trigger. The level-2 trigger algorithm was implemented in the Barrel Electromagnetic
Calorimeter (BEMC) and optimized based on the information of the direct $\gamma/\pi^0$ ratio in $Au+Au$ collisions [14], the
$\pi^0$ decay kinematics, and the electromagnetic shower profile characteristics. The BEMC has full azimuthal coverage and
pseudorapidity coverage $\mid\eta\mid$ $\leq$ 1.0. As a reference measurement we use $p+p$ data at $\sqrt{s_{NN}}$ = 200 GeV
taken in 2006 with integrated luminosity of 11 $\mathrm{pb}^{-1}$. The Time Projection Chamber (TPC) was used
to detect charged particle tracks and measure their momenta. The charged track quality cuts are similar
to previous STAR analyses [15]. For this analysis, events with at least one cluster with $E_T >$ 8~GeV were selected. 
To ensure the purity of the photon-triggered sample, trigger towers were rejected if a track with $p >$ 3~GeV/$c$ points to it.
 
A crucial step of the analysis is to discriminate between showers of direct $\gamma$ and two close $\gamma$'s from a
high-p$_{T}$ symmetric $\pi^{0}$ decay. At p$_T$ $\sim$ 8 GeV/c the angular separation between the two photons 
resulting from a symmetric $\pi^{0}$ decay (both decays photons have similar energy, smallest opening angle) at
the BEMC face is typically smaller than the tower size ($\Delta\eta=0.05,\Delta\phi=0.05$); but 
a $\pi^{0}$ shower is generally broader than a single $\gamma$ shower. The Barrel Shower Maximum Detector (BSMD), 
which resides at $\sim$ 5X$_{0}$ inside the calorimeter towers, is well-suited for 
$(2\gamma$)/$(1\gamma)$ separation up to p$_T$ $\sim$ 26 GeV/c due to its fine segmentation ($\Delta\eta\approx 0.007,\Delta\phi\approx 0.007$). 
In this analysis the $\pi^{0}$/$\gamma$ discrimination was carried out by making cuts on the 
shower shape as measured by the BSMD, where the $\pi^{0}$ identification 
cut is adjusted in order to obtain a very pure sample of $\pi^{0}$ and a sample rich in direct $\gamma$ ($\gamma_{rich}$). 
The discrimination cuts are varied to determine the systematic uncertainties. To determine 
the combinatorial background level the relative azimuthal angular distribution of the associated 
particles with respect to the trigger particle 
is fitted with two guassian peaks and a straight line.  
The near- and away-side yields, Y$^{n}$ and Y$^{a}$, of associated particles per trigger are extracted by 
integrating the $\rm 1/N_{trig} \rm dN/{\rm d}(\Delta\phi)$  
distributions in $\mid\Delta\phi\mid$ $\leq$  0.63 and $\mid\Delta\phi -\pi\mid$  $\leq$  0.63 respectively. 
The yield is corrected for the tracking efficiency of associated charged particles as a function of multiplicity. 

The shower shape cuts used to select a sample of direct photon``rich" triggers reject most of the $\pi^{0}$'s, but do not reject 
photons from highly asymmetric $\pi^{0}$ decays, $\eta$'s, and fragmentation photons. 
All of these sources of background are removed as follows from Eq.(1) below, but only within the systematic
uncertainty on the assumption that their correlations are similar to those for $\pi^{0}$'s.
Assuming zero near-side yield for direct photon triggers and a very pure sample of $\pi^{0}$, 
the away-side yield of hadrons correlated with the direct photon is extracted as
\begin{eqnarray}
Y_{\gamma_{direct}+h}=\frac{Y^{a}_{\gamma_{rich}+h}-R Y^{a}_{\pi^{0}+h}}{1-R},\label{eq2}
\hspace{0.5cm}                            
   \hspace{0.5cm}   R=\frac{Y^{n}_{\gamma_{rich}+h}}{Y^{n}_{\pi^{0}+h}}.
\end{eqnarray}
Where Y$^{a(n)}_{\gamma_{rich}+h}$ and Y$^{a(n)}_{\pi^{0}+h}$ are the away (near)-side yields of associated particles 
per $\gamma_{rich}$ and $\pi^{0}$ triggers
respectively, so that R is the fraction of $\gamma_{rich}$ triggers that are actually from $\pi^{0}$, $\eta$, and fragmentation
photons.
\section{Results}
\label{results}

\begin{figure}
\begin{center}
\begin{tabular}{cc}
   \resizebox{90mm}{153pt}{\includegraphics{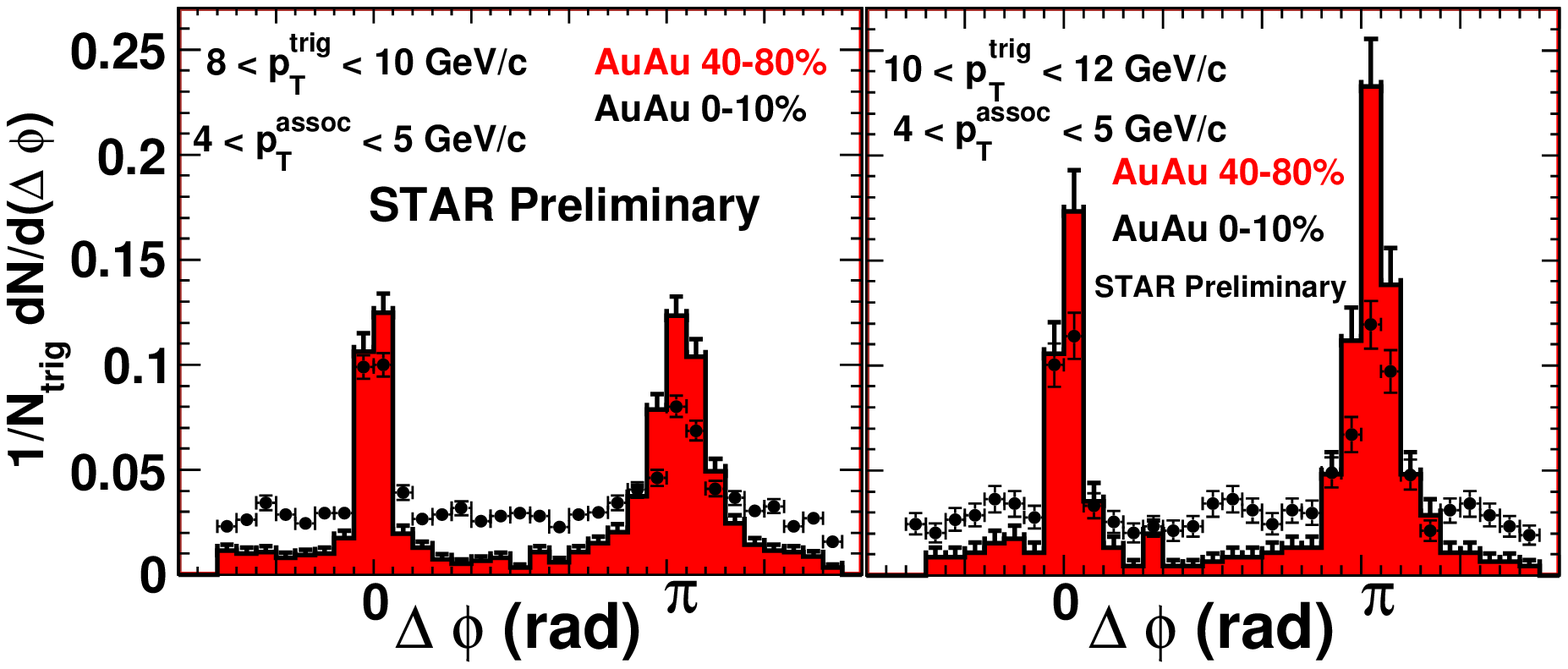}} 
    \resizebox{90mm}{160pt}{\includegraphics{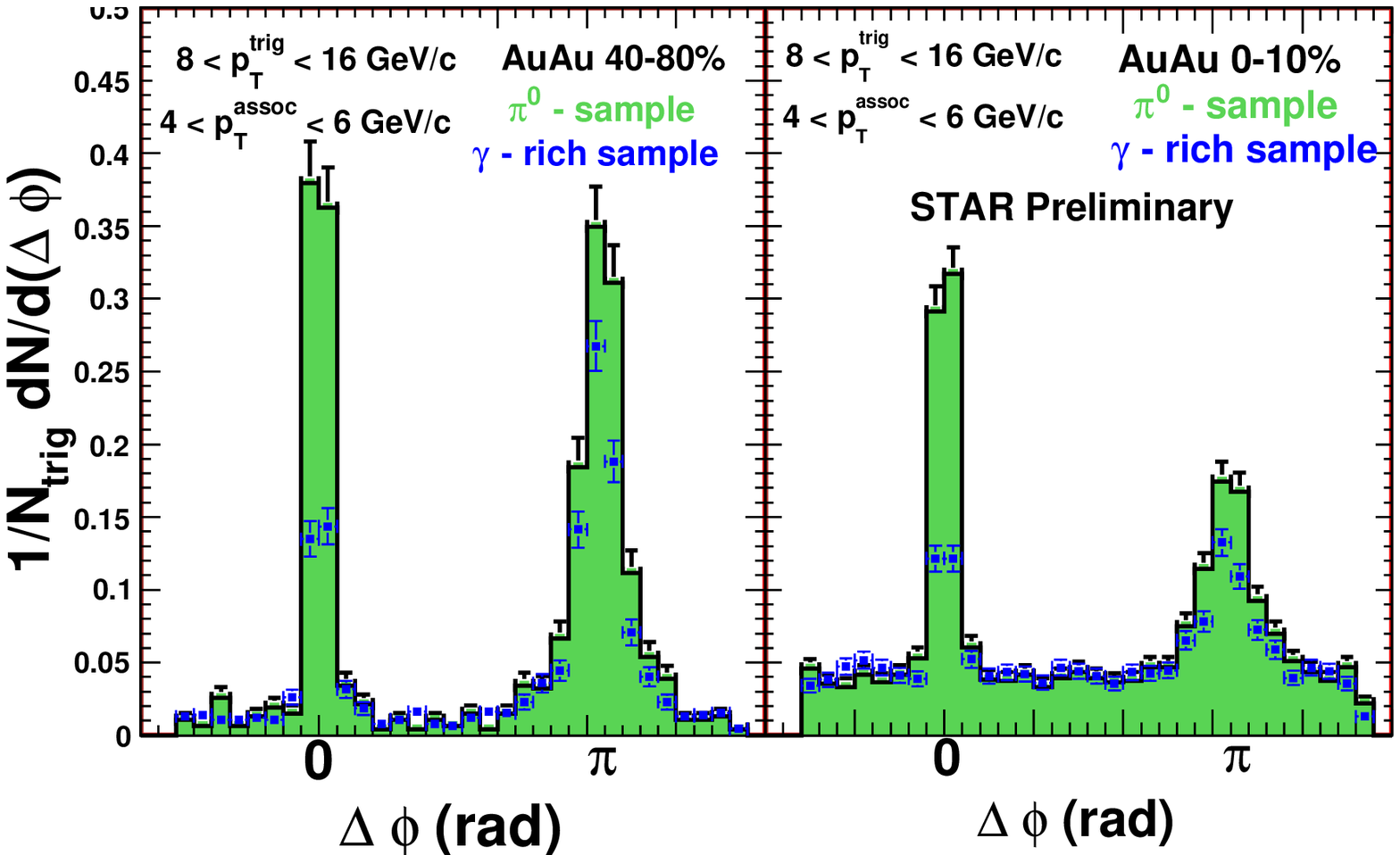}} \\
        \end{tabular}
    \caption{Left: Azimuthal correlation histograms of high-p$_{T}^{trig}$ inclusive photons 
    with associated hadrons for 40-80$\%$ and 0-10$\%$ $Au+Au$
    collisions. Right: Azimuthal correlation histograms of high-p$_{T}^{trig}$ $\gamma$-rich sample and $\pi^{0}$-sample  
    with associated hadrons for 40-80$\%$ and 0-10$\%$ $Au+Au$
    collisions}
      \end{center} 
      \label{fig:1} 
\end{figure}
Figure 1 (left) shows the azimuthal correlation for inclusive photon triggers for the most peripheral and central bins in Au+Au collisions. 
Parton energy loss in the medium causes the away-side to be increasingly suppressed with centrality as it was previously reported
[3,15].
The suppression of the near-side yield with centrality, which has not been observed in the charged hadron 
azimuthal correlation, is consistent with an increase of the $\gamma$/$\pi^{0}$ ratio with centrality at high E$_{T}^{trig}$. 
The shower shape analysis is used to distinguish between the $(2\gamma$)/$(1\gamma)$ showers as in Figure 1 (right) 
which shows the azimuthal correlation for $\gamma$-rich sample triggers and $\pi^{0}$ triggers for the
most peripheral and central bins. The $\gamma$-rich sample has a lower near-side yield than $\pi^{0}$-triggered sample, but it is not zero.
The non-zero near-side yield for the $\gamma$-rich sample is expected due to the remaining contributions of the widely 
separated photons from other
sources, because the shower shape analysis is only effective for the two close $\gamma$ showers. 

The purity of $\pi^{0}$
identification with the shower shape analysis is verified by comparing to previous measurements of azimuthal correlations between charged
hadrons ($ch-ch$) [15]. 
\begin{figure}
\begin{center}
\begin{tabular}{cc}
   \resizebox{120mm}{!}{\includegraphics{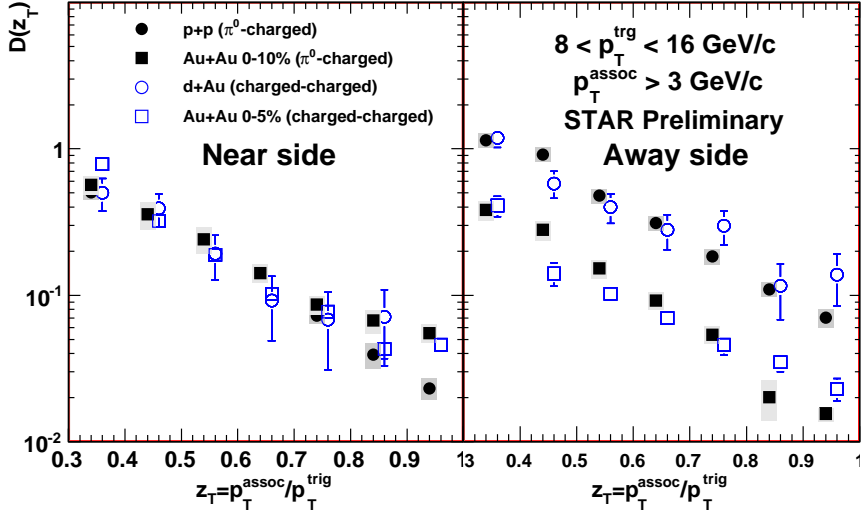}} 
        \end{tabular}
    \caption{$z_{T}$ dependence of $\pi^{0}$-ch and $ch-ch$ [15] near-side (left panel) and away-side (right panel) associated particle yields.}
      \end{center} 
      \label{fig:2} 
\end{figure}
Figure 2 shows the $z_{T}$ dependence of the associated hadron yield normalized per $\pi^{0}$ trigger D($z_{T}$), where $z_{T}= p_{T}^{assoc}/p_{T}^{trig}$
[16], for the near-side and away-side compared to
the per charged hadron trigger [15]. The near-side yield as in
Figure 2 (left) shows no significant difference between $p+p$, $d+Au$, and $Au+Au$ indicating in-vacuum fragmentation even in 
heavy-ion collisions, This can be due to either a
surface bias as generated in several model calculations [17,18,19,20] or the parton fragmenting in vacuum after losing energy in the
medium. However the medium effect is clearly 
seen in the away-side in Figure 2 (right) where the per trigger yield in
$Au+Au$ is significantly suppressed compared to $p+p$ and $d+Au$. The general agreement between 
the results from this analysis
($\pi^{0}$-$ch$) and the previous analysis ($ch-ch$) is clearly seen in both panels of Figure 2 
which indicates the
purity of the $\pi^{0}$ sample and therefore the effectiveness of the shower shape cut to identify $\pi^{0}$.

The away-side associated yields per trigger photon for direct $\gamma$-charged hadron correlations are extracted using Eq. 1.  
Figure 3 (left) shows the $z_{T}$ dependence of the trigger-normalized fragmentation function for $\pi^{0}$-charged correlations 
($\pi^{0}$-ch)
compared to measurements with direct $\gamma$-charged correlations ($\gamma$-$ch$). The away-side yield per 
trigger of direct 
$\gamma$ is smaller than with $\pi^{0}$ trigger at the same centrality class. This difference is due to 
the fact that the 
$\pi^{0}$ originates from higher initial parton energy and therefore has a larger associated jet multiplicity. 
\begin{figure}
 \begin{center}
  \begin{tabular}{cc}
   \resizebox{90mm}{!}{\includegraphics{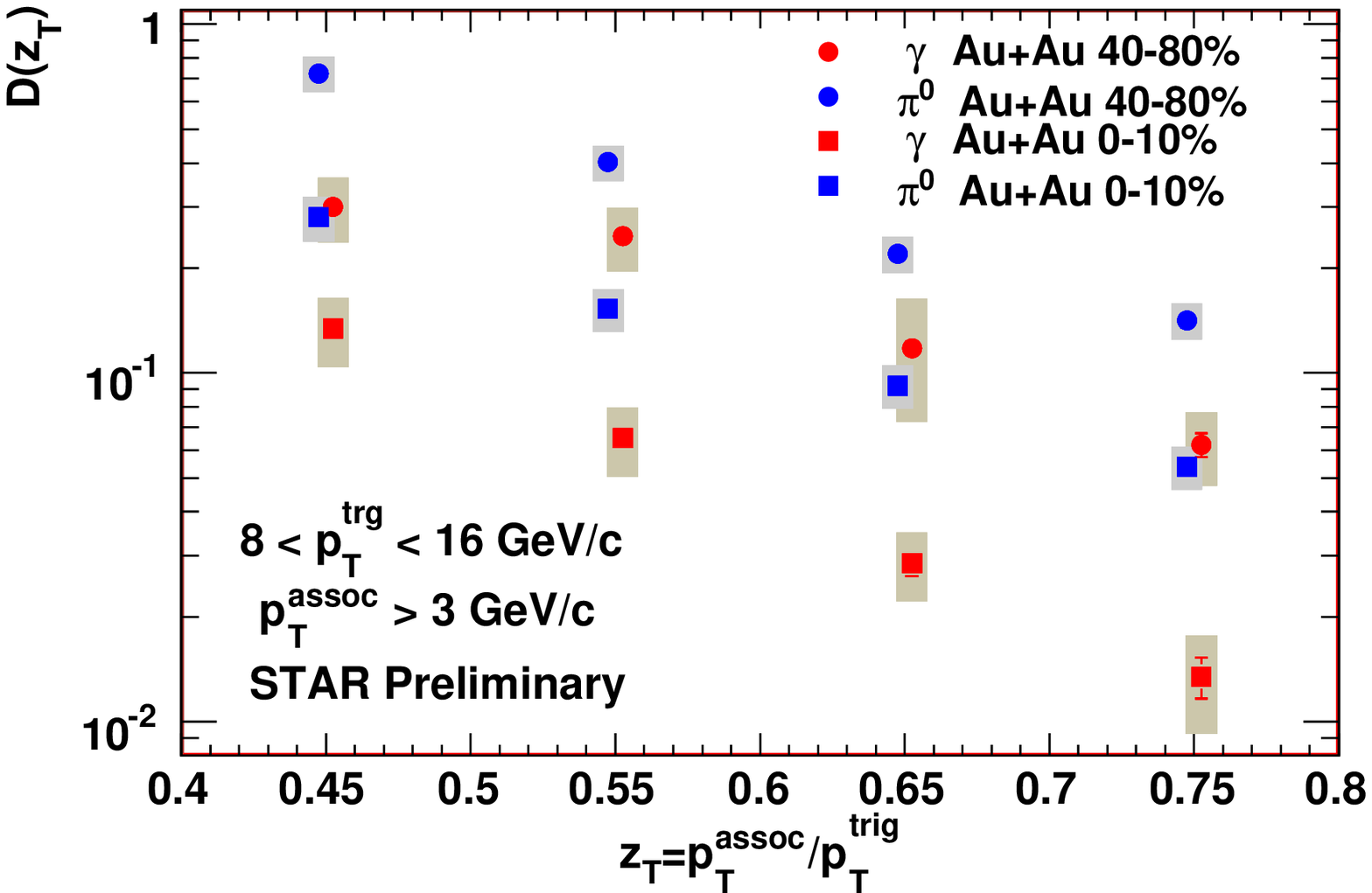}} 
   \resizebox{90mm}{!}{\includegraphics{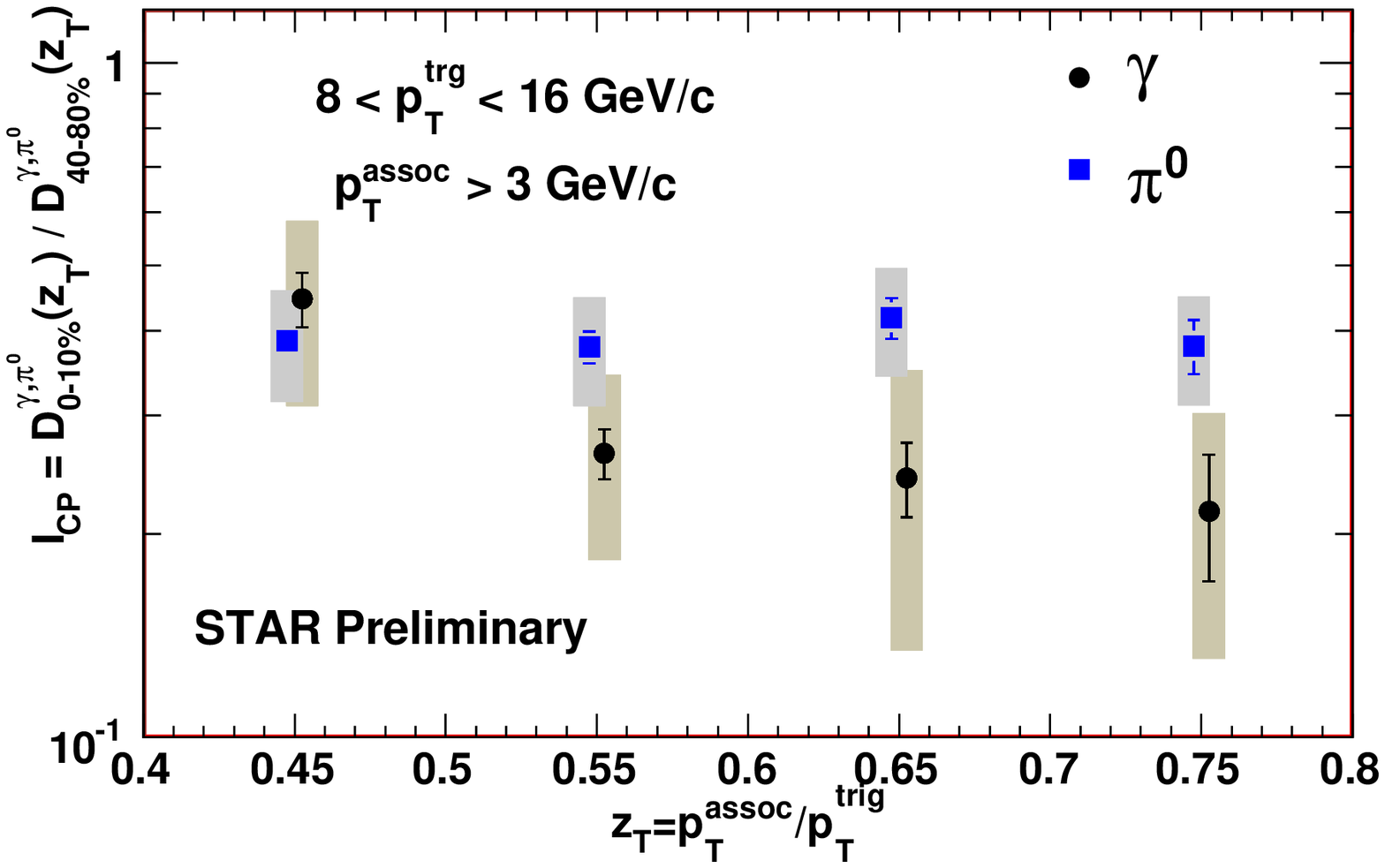}} \\
     \end{tabular}
    \caption{(Left) $z_{T}$ dependence of associated recoil yield with $\pi^{0}$ and direct $\gamma$ triggers for 40-80$\%$ and
    0-10$\%$ $Au+Au$ collisions. (Right) $z_{T}$ dependence of I$_{CP}$ for direct $\gamma$ triggers and $\pi^{0}$ triggers (see text). Boxes
    show the systematic uncertainties.}
    \end{center}
    \label{fig:3}
\end{figure}
\begin{figure}
 \begin{center}
  \begin{tabular}{cc}
   \resizebox{90mm}{!}{\includegraphics{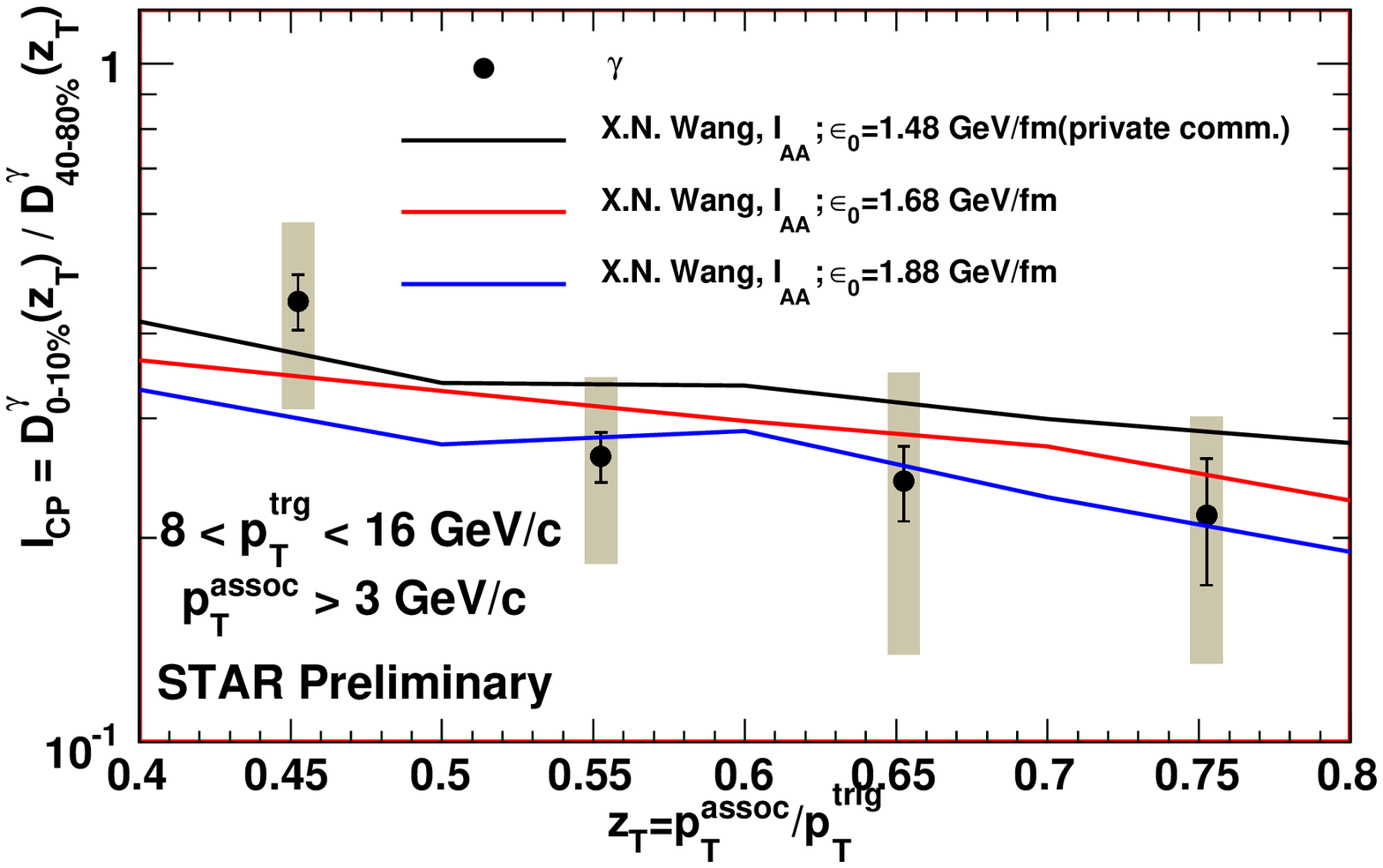}} 
   \resizebox{90mm}{!}{\includegraphics{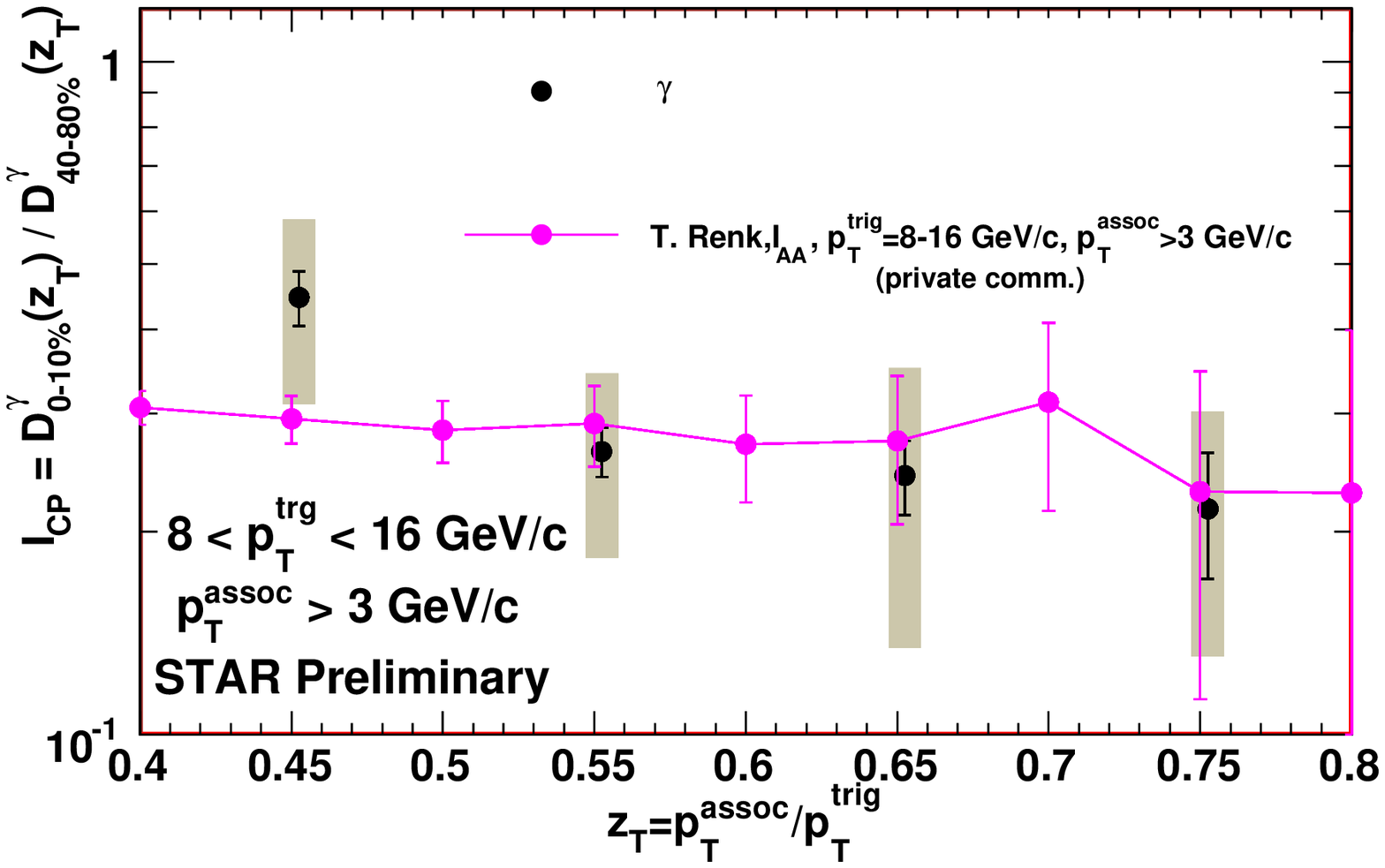}} \\
     \end{tabular}
    \caption{$z_{T}$ dependence of $I_{CP}$ for direct $\gamma$ triggers associated particle yields compared with theoretical calculations
    (left) $I_{AA}$ of 0-10$\%$ $Au+Au$ collisions (only Annihilation and Compton processes to NLO are 
    considered in the theoretical calculations) with three different initial gluon density where 7 $<$ p$_T^{trig}$ $<$ 9 GeV and p$_T^{assoc}$ $>$ 5 GeV/c, (Right) $I_{AA}$ where 
    8 $<$ p$_T^{trig}$ $<$ 16 GeV and p$_T^{assoc}$ $>$ 3 GeV/c.}
    \end{center}
    \label{fig:4}
\end{figure}

In order to quantify the away-side suppression, we calculate the quantity I$_{CP}$, which is defined as the 
ratio of the integrated yield of 
the away-side associated particles per trigger particle in $Au+Au$ central (0-10$\%$ of the geometrical cross section) 
relative to $Au+Au$
peripheral (40-80$\%$ of the geometrical cross section) collisions.
Figure 3 (right) shows the I$_{CP}$ for $\pi^{0}$ triggers and for direct $\gamma$ triggers 
as a function of $z_{T}$. The ratio would be unity if there were no medium effects on the parton fragmentation; 
the observed ratio deviates
from unity by a factor of $\sim$ 2.5. The ratio for the $\pi^{0}$ trigger is approximately independent of $z_{T}$ for the shown
range in agreement with the previous results from ($ch-ch$) measurements [15]. 
Within the current systematic uncertainty the I$_{CP}$ of direct $\gamma$ and $\pi^{0}$ are similar.    

Suppression ratios with respect to the p+p reference, I$_{AA}$, have
been reported earlier [21]. The values of I$_{AA}$ are smaller than for I$_{CP}$,
indicating finite suppression in the peripheral 40-80$\%$ data, but the
statistical uncertainties are large due to the small $\gamma$/$\pi^{0}$ ratio in p+p as previously reported [22]. Nevertheless, 
the value of I$_{AA}$ is found to be similar to the values observed
for di-hadron correlations and for single-particle suppression R$_{AA}$.  

A comparison of I$_{CP}$ of direct $\gamma$-triggered yields with two theoretical model calculations of I$_{AA}$ 
is shown in Figure 4.
The I$_{CP}$ values agree well with the theoretical predictions within the current uncertainties.
Figure 4 (left) indicates that a reduction in the systematic and statistical uncertainties is needed 
to distinguish between different color charge densities.   

\section{Summary and Outlook}
\label{summary and outlook}
In summary, the first measurement of fragment distributions for jets with energy controlled via $\gamma$-jet in $Au+Au$ collisions has been performed by the 
STAR experiment. The STAR detector is unique to perform such correlation measurements due to the full coverage in azimuth. 
Within the current uncertainty the recoil suppression ratio I$_{CP}$ of direct $\gamma$ and $\pi^{0}$ are similar.
A full analysis of the systematic uncertainties is under way and may
lead to a reduction of the total uncertainty. Future RHIC runs will
provide larger data samples to further reduce the uncertainties and
extend the $z_{T}$ range.

\end{document}